\documentclass{iaus}
\usepackage{graphicx}

\title[Extrasolar terrestrial planets] 
{The Diversity of Extrasolar Terrestrial Planets.}

\author[J. C. Bond, D. S. Lauretta \& D. P. O'Brien]   
{Jade C. Bond$^1$, Dante S. Lauretta$^2$
 \and David P. O'Brien$^1$}

\affiliation{$^1$Planetry Science Institute, \\ 1700 E Fort Lowell Rd, Tucson AZ 85719 \\ email: {\tt jbond@psi.edu} \\[\affilskip]
$^2$Lunar and Planetary Laboratory, University of Arizona, \\ 1629 E. University Blvd., Tucson AZ 85721}

\pubyear{2009}
\volume{265}  
\pagerange{119--126}
\jname{Chemical Abundances in the Universe: Connecting First Stars to Planets}
\editors{K. Cunha, M. Spite \& B. Barbuy, eds.}
\begin{document}

\maketitle

\begin{abstract}
Extrasolar planetary host stars are enriched in key planet-building elements. These enrichments have the potential to drastically alter the building blocks available for terrestrial planet formation. Here we report on the combination of dynamical models of late-stage terrestrial planet formation within known extrasolar planetary systems with chemical equilibrium models of the composition of solid material within the disk. This allows us to constrain the bulk elemental composition of extrasolar terrestrial planets. A wide variety of resulting planetary compositions exist, ranging from those that are essentially "Earth-like", containing metallic Fe and Mg-silicates, to those that are dominated by graphite and SiC. This implies that a diverse range of terrestrial planets are likely to exist within extrasolar planetary systems.
\end{abstract}

\firstsection
\section{Introduction}
To date, more than 400 extrasolar giant planets are known to exist, the vast majority being Neptune-mass or larger. The recent discovery of super-Earth-sized extrasolar terrestrial planets suggests that Earth-mass terrestrial planets may also be present within many of these systems. Given that the number of planets in the galaxy is expected to correlate inversely with planetary mass, it is expected that Earth-sized terrestrial planets are much more common than giant planets (\cite{marcy:00}). The possibility of terrestrial extrasolar planets has been examined by several authors, with many focussing on the long-term dynamical stability of terrestrial planet forming regions (e.g. \cite{br2}). Others have considered terrestrial planet formation in hypothetical systems (\cite{ray05,avi}) while only one study has examined terrestrial planet formation within a specific planetary system (\cite{br3}). While such studies are undoubtably of great benefit to future planet-finding missions, they do not provide any insight into the potential nature of a given terrestrial planet.

At the same time, extrasolar planetary host stars are chemically distinct from the general stellar population, displaying significant variations in key planet-building elements such as Fe, C, O, Mg and Si (\cite{bond:2008} and references therein). These enrichments are primordial in origin, established in the giant molecular cloud from which these systems formed (e.g. \cite{bond:2008}). As a result, the chemistry of planet-building materials in many of these systems is distinctly different from that in our Solar System. Consequently, it is likely that terrestrial extrasolar planets may have compositions reflecting the enrichments observed in the host stars, possibly resulting in terrestrial planets with compositions and mineralogies unlike any body yet observed within our Solar System.

\section{Planetary Building Blocks}
Two of the most important elemental ratios for determining the mineralogy of extrasolar terrestrial planets are C/O and Mg/Si. The C/O ratio controls the distribution of Si among carbide and oxide species. If the C/O ratio is greater than 0.8, Si exists in solid form primarily as SiC. Additionally, significant amounts of solid C are present as planet building materials. For C/O values below 0.8, Si is present in rock-forming minerals as the SiO{$_2$} structural unit. Silicate mineralogy is controlled by the Mg/Si value. For Mg/Si values less than 1, Mg is in pyroxene and the excess Si is present as other silicate species such as feldspars. For Mg/Si values ranging from 1 to 2, Mg is distributed primarily between olivine and pyroxene.  For Mg/Si values extending beyond 2, all available Si is consumed to form olivine.

\begin{figure}[b]
\begin{center}
 \includegraphics[width=3.4in]{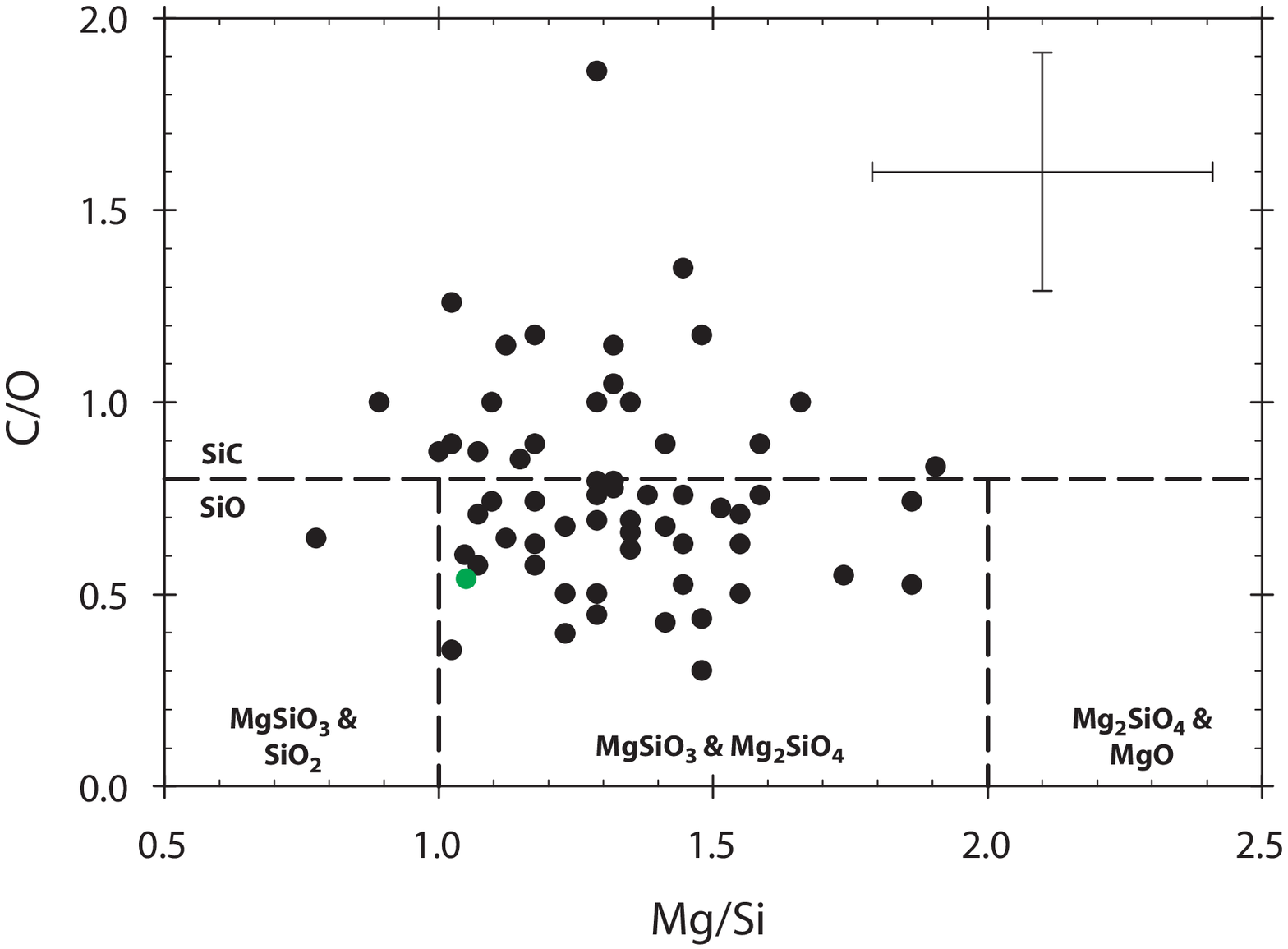}
 \caption{Mg/Si vs. C/O for known planetary host stars
with reliable stellar abundances. Stellar photospheric values were taken from \cite{gilli} (Si, Mg), \cite{be} (Mg), \cite{eca} (C) and
\cite{oxygen} (O). Solar values are shown by the green circle and were taken from \cite{asp}. The dashed line indicates a C/O value of 0.8 and
marks the transitions between a silicate-dominated composition and a carbide-dominated composition at 10$^{-4}$. Average 2-$\sigma$ error bars
shown in upper right.}
   \label{fig1}
\end{center}
\end{figure}

The bulk compositions of extrasolar planetary host stars vary greatly in their C/O and Mg/Si values (Fig. \ref{fig1}). The C/O ratios for all known extrasolar planetary systems range from 0.30 to 1.86 with a mean value of 0.77$\pm$0.31. The Mg/Si ratios show a similar variation, ranging from 0.78 to 1.91 with a mean of 1.32$\pm$0.31. It should be noted that the Solar System is borderline anomalous within this population with C/O$_{\bigodot}$ = 0.54 and Mg/Si$_{\bigodot}$ = 1.044. This compositional variation implies that a wide variety of terrestrial planet compositions are present within extrasolar planetary systems. The fact that most extrasolar planetary systems have non-solar elemental abundances suggests that the terrestrial planets may differ in composition from the Earth.

\section{Extrasolar Terrestrial Planet Formation}
In order to quantify the diversity of extrasolar terrestrial planet compositions, we selected several planetary systems spanning the entire range of observed Mg/Si and C/O values for detailed study. We determined simulated terrestrial planet compositions in the same way as was previously successfully applied to the Solar System by Bond et al (2009). Stellar photospheric abundances of the 14 most abundant elements (C, N, O, Na, Mg, P, Al, Si, S, Ca, Ti, Cr, Fe and Ni) (\cite{gilli,be,ec,oxygen}) were utilized to obtain condensation sequences. In order to simulate the final stages of planetary accretion, n-body dynamical simulations were run using the SyMBA symplectic n-body integrator (\cite{duncan}). The giant planets are assumed to be in their current positions at the beginning of the simulations, corresponding to a case where they migrated to those locations early enough that planetesimals and embryos were able to form after giant planet migration (\cite{armitage:03}).

\section{Extrasolar Terrestrial Planet Compositions}
The primary condensate mineralogy of several of our model systems is essentially identical to that of the Solar System. For example, HD72659 is very similar to the Solar System, both in terms of its bulk mineralogy and its general chemical structure (such as the pronounced appearance of the water ice line at 140K). Combination of the condensate mineralogy with the dynamical simulations shows that terrestrial planets in HD72659 have compositions similar to those in our own Solar System. Enrichments occur in the planetary abundances of the most volatile species (Na and S). This enrichment is likely an artifact of the fact that we do not consider volatile loss during accretion in the current simulations, which is expected to be significant. Thus, we can expect that terrestrial planets present within these systems have compositions broadly similar to those within our own Solar System. Therefore, systems with elemental abundances and ratios similar to these are ideal places to focus future ``Earth-like'' planet searches.

Profound mineralogical variations occur in systems with C/O values above approximately 0.8. The inner part of these systems are dominated by refractory carbon species such as graphite, SiC and TiC. Outside of this zone, the mineralogy largely resembles that of the Solar System. The presence of this narrow carbon zone results in Earth-like terrestrial bodies enriched in C (up to 0.68wt\%). The Gl777A system has Mg/Si and C/O values that are close to the mean values of known extrasolar planetary systems (1.32 and 0.78 respectively). This result implies that the ``average'' extrasolar planetary system is made of solid material with compositions comparable to that of our own Solar System with minor C enrichment.

It is the systems with the highest C/O values that produce the most unusual planets. For example, HD108874 has a C/O value of 1.35. In this system, refractory species are composed of C, SiC and TiC, as opposed to the Ca and Al-rich inclusions characteristic of the earliest solids from our Solar System. Solid carbon is stable inside of 1.4 AU, causing close-in terrestrial planets to be dominated by carbon-bearing solids. For example, HD108874 may contain terrestrial planets that are dominated by C and SiC, containing up to 68 wt\% C with smaller amounts of Si and Fe.

Of the 60 systems with reliable O abundances, 21 have C/O values above 0.8, implying that carbide minerals are important planet building materials in approximately 35\% of planetary systems. Given the uncertainties on the stellar elemental abundances themselves,  this value is an upper limit on the prevalence of C-rich systems. We can say, however,  with 2$\sigma$ confidence that at least 6 known planetary systems have C/O values above 0.8 (10\% of the sample) and would thus contain C and its associated phases as a major planet forming material. These data clearly demonstrate that there are a significant number of systems in which terrestrial planets have compositions vastly different to any planetary body observed in our Solar System. These bodies may represent a new class of terrestrial planet.

The distribution of water vapor is also altered within the C-rich systems. Instead of being present throughout the entire disk (as for systems with lower C/O values), water vapor is only present in significant quantities at temperatures below 600 - 800K (generally outside of 0.8 AU). Currently, one theory regarding the delivery of water to the early Earth is that it occurred locally via adsorption onto solid grains that were later accreted onto the Earth (\cite{drake:06}). Thus the restricted distribution of water vapor may prohibit the formation of ocean-bearing planets in these systems.

Finally, the mass distribution of these high-C/O systems is intriguing. The combination of a broad zone of refractory carbon-bearing solids and the relatively small amount of water ice that condenses in these systems suggests that these systems may have more solid mass located in the inner regions (inside 1 AU) of the disk than for Solar-like disks. This mass distribution is the opposite of that expected in O-rich systems and suggests that protoplanets may form more easily in the inner regions of these systems. The full implications of this need to be examined by using alternative mass distributions for extrasolar planetary formation simulations for both gas giant planets and smaller terrestrial planets.

\section{Implications and Ongoing Work}
The results of our study clearly have a wide variety of implications ranging from planetary processes (such as plate tectonics and atmospheric reactions) to planetary detectability and even astrobiology and are the subject of ongoing research. As giant planet migration almost certainly occurred in a large number of known planetary systems, simulations incorporating migration are underway. However, the current in-situ simulations clearly indicate that a truly diverse range of terrestrial extrasolar planets are likely to exist.


\begin{thebibliography}{}

\bibitem[Armitage (2003)]{armitage:03}
{Armitage, P. J.} 2003,
\textit{ApJ}, 582, L47

\bibitem[Asplund \etal\ (2006)]{asp}
{Asplund, M., Grevesse, N. \& Sauval, A.J.} 2005,
\textit{ASPC}, 336, 25

\bibitem[Beir{\~a}o \etal\ (2005)]{be}
{Beir{\~a}o, P., Santos, N.~C., Israelian, G. \& Mayor, M.} 2005,
\textit{A\&A}, 438, 251

\bibitem[Bond \etal\ (2008)]{bond:2008}
{Bond, J.C. et. al.} 2008,
\textit{ApJ}, 682, 1234

\bibitem[Bond \etal\ (2009)]{bond:09}
{Bond, J. C., O'Brien, D. P \& Lauretta, D. S.} 2009,
\textit{Icarus}, In press.

\bibitem[Drake \& Campins (2006)]{drake:06}
{Drake, M. J. \& Campins, H.} 2006,
\textit{Proc. IAU}, 381

\bibitem[Ecuvillon \etal\ (2004)]{eca}
{Ecuvillon, A. et al.} 2004,
\textit{A\&A}, 426, 619

\bibitem[Ecuvillon \etal\ (2004)]{ec}
{Ecuvillon, A. et al.} 2004,
\textit{A\&A}, 426, 629

\bibitem[Ecuvillon \etal\ (2006)]{oxygen}
{Ecuvillon, A. et al.} 2006,
\textit{A\&A}, 445, 633

\bibitem[Duncan \etal\ (2005)]{duncan}
{Duncan, M.J., Levison, H.F. \& Lee, M.H.} 1998,
\textit{AJ}, 116, 2067

\bibitem[Gilli \etal\ (2006)]{gilli}
{Gilli, G., Israelian, G., Ecuvillon, A., Santos, N.~C. \& Mayor, M.} 2006,
\textit{A\&A}, 449, 723

\bibitem[Mandell \etal\ (2007)]{avi}
{Mandell, A.M., Raymond, S.N. \& Sigurdsson, S.} 2007,
\textit{ApJ}, 660, 823

\bibitem[Marcy \& Butler (2000)]{marcy:00}
{Marcy, G. W. \& Butler, R. P.} 2000,
\textit{ASPC}, 213

\bibitem[Raymond \& Barnes (2005)]{br2}
{Raymond, S.N. \& Barnes, R.} 2005,
\textit{ApJ}, 619, 549

\bibitem[Raymond \etal\ (2005)]{ray05}
{Raymond, S.N., Quinn, T. \& Lunine, J.I.} 2005,
\textit{Icarus}, 177, 256

\bibitem[Raymond \etal\ (2006)]{br3}
{Raymond, S.N., Barnes, R. \& Kaib, N. A.} 2006,
\textit{ApJ}, 644, 1223

\end{thebibliography}
\end{document}